\documentclass[twocolumn,aps,showpacs,amsmath,amssymb,superscriptaddress]{revtex4}
\usepackage{graphicx}
\usepackage{dcolumn}
\usepackage{bm}

\def\beqa{\begin{eqnarray}}
\def\enqa{\end{eqnarray}}
\def\beq{\begin{equation}}
\def\enq{\end{equation}}
\newcommand{\nc}{\newcommand}
\nc{\rnc}{\renewcommand}
\nc{\nn}{\nonumber}
\nc{\ch}{\cosh}
\nc{\sh}{\sinh}
\rnc{\th}{\tanh}
\nc{\db}{\displaybreak[0]\\}
\nc{\bra}{\langle}
\nc{\ket}{\rangle}
\rnc{\(}{\left(}
\rnc{\)}{\right)}
\nc{\xxx}{$XXX \,$}
\nc{\lam}{\lambda}
\begin{document}
\title{Next Nearest-Neighbor Correlation Functions of the Spin-1/2 XXZ 
Chain at Massive Region}
\author{Minoru Takahashi}
\email[]{mtaka@issp.u-tokyo.ac.jp}
\affiliation{Institute for Solid State Physics, University of Tokyo, Kashiwa,
Chiba 277-8571, Japan}
\author{Go Kato}
\email[]{kato@monet.phys.s.u-tokyo.ac.jp}
\affiliation{Department of Physics, Graduate School of Science, University of 
Tokyo, Hongo 7-3-1, Bunkyo-ku, Tokyo 113-0033, Japan}
\author{Masahiro Shiroishi}
\email[]{siroisi@issp.u-tokyo.ac.jp}
\affiliation{Institute for Solid State Physics, University of Tokyo, Kashiwa,
Chiba 277-8571, Japan}
%
\begin{abstract}
The second neighbor correlation functions of the spin-${\frac{1}{2}}$ $XXZ$ chain 
in the ground state are expressed in the form of three dimensional integrals.
We show that these integrals can be reduced 
to one-dimensional ones and thereby evaluate the values of the 
next nearest-neighbor correlation functions for ${\Delta >1}$. 
\end{abstract}

\pacs{75.10.Jm, 75.50.Ee, 02.30.Ik}
\maketitle
%
%
\section{Introduction}
The spin-1/2 ${XXZ}$ chain is one of the fundamental models in the study 
of the low-dimensional magnetism. The Hamiltonian is given by
\begin{align}
H=\sum_{j=1}^{N} 
\left\{ S^x_j S^x_{j+1} + S^y_j S^y_{j+1} 
+ \Delta S^z_j S^z_{j+1} \right\}, \label{XXZ}
\end{align}
where  $S= \sigma /2$ and  $\sigma$ are Pauli matrices.
The model can be solved by Bethe ansatz method and diverse physical properties 
have been investigated with varying anisotropy parameter $\Delta$ 
\cite{Bethe31,Hulthen38,TakahashiBook}.  For $\Delta=1$ we know the second neighbor 
correlation \cite{Takahashi77}.  Jimbo et al \cite{Jimbo92,Nakayashiki94,Jimbo96} 
obtained the multiple integral representation 
for the arbitrary correlation functions of the ${XXZ}$ chain for ${1\ge \Delta> -1}$ and
$\Delta>1$. Later Kitanine, Maillet and Terras\cite{Kitanine00} 
generalized to magnetized case using the Bethe wave function. 
But this multiple integral is very complicated and they only reproduce the nearest neighbor correlations. 
Recently Boos et al 
succeeded to calculate these multiple integrals for emptiness formation probability  (EFP) 
\beq
P(m)=\langle \prod_{j=1}^m(S_j^z+{1\over 2})\rangle,\label{Pm}
\enq 
at $\Delta=1, m=3,4,5,6$\cite{Boos01,Boos02,BKNS02,Boos03}. 
The case $m=3$ reproduces the result of the second neighbor correlation. The same method is applied to the 
third neighbor correlations at $\Delta=1$\cite{Sakai03}. 
In the previous paper we calculated the second neighbor corelation functions for massless
case $-1<\Delta\le 1$ and integral is simplified to the one dimensional integral\cite{Kato03}. 
In this paper we give the second neighbor correlation 
in massive region ${1 < \Delta }$. The correlation formula 
contains the elliptic functions. But we can show that the final result also contain only one-dimensional integral.

\begin{widetext}
\section{Multiple integral formula}
\subsection{Massless case}
The correlation function in massless case is given by,
\begin{align}
&F\Bigl({ 
\begin{matrix}
{\epsilon_1,\epsilon_2,...,\epsilon_m}\cr {\epsilon_1',\epsilon_2',...,\epsilon_m'}
\end{matrix}}\Bigl)
=\frac{\langle \psi_g |E^{\epsilon_1}_{\epsilon'_1}E^{\epsilon_2}_{\epsilon'_2}...
E^{\epsilon_m}_{\epsilon'_m}|\psi_g\rangle}{\langle \psi_g |\psi_g\rangle}=
(-1)^s(-{\pi\over \zeta})^{m(m+1)\over 2}
\prod^{s'}_{j=1}\int^{\infty-i\zeta}_{-\infty-i\zeta}{d\lambda_j\over 2\pi}
\prod^m_{j=s'+1}\int^\infty_{-\infty}{d\lambda_j\over 2\pi}
\prod_{a>b}{\sinh{\pi\over\zeta}(\lambda_a-\lambda_b)\over
\sinh(\lambda_a-\lambda_b-i\zeta)}
\nonumber\\
&\times\prod_{j\in \alpha^+}{\sinh^{j-1}(\mu_j'+3i{\zeta\over 2})
\sinh^{m-j}(\mu_j'+i{\zeta\over 2})\over(\cosh{\pi\over\zeta}\mu_j')^m}
\prod_{j\in \alpha^-}{\sinh^{j-1}(\mu_j-i{\zeta\over 2})
\sinh^{m-j}(\mu_j+i{\zeta\over 2})\over(\cosh{\pi\over\zeta}\mu_j)^m},\label{mless}
\end{align}
where
\begin{align}
&\Delta=\cos \zeta,\quad\{\lambda_1,...,\lambda_m\}=\{\mu'_{j'_{max}},...,\mu'_{j'_{min}},
\mu_{j_{min}},...,\mu_{j_{max}}\},\nonumber\\
&\alpha^+=\{j;1\le j\le m, \epsilon_j=+\}, {\rm card}(\alpha^+)=s',\quad
{\rm Max}_{j\in \alpha^+}(j)\equiv j'_{max}, {\rm Min}_{j\in\alpha^+}(j)=j'_{min},\nonumber\\
&\alpha^-=\{j;1\le j\le m, \epsilon_j'=-\}, {\rm card}(\alpha^-)=s,\quad 
{\rm Max}_{j\in \alpha^-}(j)\equiv j_{max}, {\rm Min}_{j\in\alpha^-}(j)=j_{min}.
\end{align}
\end{widetext}
\begin{widetext}
If we put $\lambda_j\to\lambda_j-i\zeta$ for $j=1,...,s'$, these equations are written as follows
\begin{align}
&F\Bigl({
\begin{matrix}
{\epsilon_1,\epsilon_2,...,\epsilon_m}\cr {\epsilon_1',\epsilon_2',...,\epsilon_m'}
\end{matrix}}
\Bigl)=
(-1)^m(-{\pi\over \zeta})^{m(m+1)\over 2}\nonumber\\
&\times\prod^m_{j=1}\int^{\infty}_{-\infty}\frac{{\rm d}\lambda_j}{2\pi}
{\displaystyle\prod_{a>b} \sinh{\pi\over \zeta}(\lambda_a-\lambda_b)\over\displaystyle
\prod^m_{j=1}\cosh^m{\pi\lambda_j\over\zeta}}{\displaystyle\prod_{j=1}^m\sinh^{y_j-1}(\lambda_j
+i{\zeta\over 2})
\sinh^{m-y_j}(\lambda_j-i{\zeta\over 2})
\over\displaystyle\prod_{a>b}
\sinh(\lambda_a-\lambda_b-i f_{ab}\zeta )},
\end{align}
where
\begin{align}
&f_{ab}=(1+{\rm sign}[(s'+1/2-a)(s'+1/2-b)])/2,\nonumber\\
&y_1>y_2,...>y_{s'},  \quad\epsilon_{y_i}=+, 
\quad y_{s'+1}>y_{s'+2}>...>y_m,\quad\epsilon'_{m+1-y_i}=-.\label{fyj}
\end{align}
If we put $\lambda_j\to x_j\zeta$ this is written as 
\begin{align}
\prod^m_{j=1}\int^{\infty}_{-\infty}{\rm d}x_j
{ W}_m(x_1,...,x_m)
{\displaystyle\prod_{j=1}^m\sinh^{y_j-1}\zeta(x_j+{i\over 2})
\sinh^{m-y_j}\zeta(x_j-{i\over 2})
\over\displaystyle\zeta^{m(m-1)/2}\prod_{a>b}
\sinh\zeta(x_a-x_b-i f_{ab})}.
\end{align}
The weight function of the integral is common for all the correlation 
functions:
\begin{align}
W_m(x_1,...,x_m)={\displaystyle\prod_{a>b}(-\pi)\sinh\pi(x_a-x_b)
\over\displaystyle\prod_{j=1}^m 2\cosh^m\pi x_j}=D_m(\prod_{j=1}^m \frac{1}{2}{\rm sech}(\pi x_j))
=D_m(\prod_{j=1}^m W_1(x_j)),
\end{align}
where $D_m$ is antisymmetric differential operator defined by
\begin{align}
D_m={1\over \prod_{r=1}^{m-1} r!}
\prod_{j>k}({\partial\over\partial x_j}-{\partial\over\partial x_k}).
\end{align}
Especially in the limit $\zeta\to 0$ or $\Delta=1$ we have 
\begin{align}
\prod^m_{j=1}\int^{\infty}_{-\infty}{\rm d}x_j
{ W}_m(x_1,...,x_m)
{\displaystyle\prod_{j=1}^m(x_j+{i\over 2})^{y_j-1}
(x_j-{i\over 2})^{m-y_j}
\over\displaystyle\prod_{a>b}(x_a-x_b-i f_{ab})}.
\end{align}
These integrals are done analytically up to $m=4$ for $\Delta=1$ \cite{Sakai03}, and up to $m=3$ for 
$-1<\Delta<1$\cite{Kato03}.

\subsection{Massive case}
For massive case integral formula for correlation function contains elliptic 
functions\cite{Jimbo92,Jimbo96,Kitanine00} 
{
\begin{align}
 &F\Bigl({
\begin{matrix}
{\epsilon_1,\epsilon_2,...,\epsilon_m}\cr {\epsilon_1',\epsilon_2',...,\epsilon_m'}
\end{matrix}}
\Bigl)=(i)^{s-s'}C_m
\prod^{s'}_{j=1}\int^{\pi/2-i\phi}_{-\pi/2-i\phi}{d\lambda_j\over 2\pi}
\prod^m_{j=s'+1}\int^{\pi/2}_{-\pi/2}{d\lambda_j\over 2\pi}\vartheta_2(\sum_{j=1}^m(\lambda_j+i\phi/2))
\prod_{a>b}{ \vartheta_1(\lambda_a-\lambda_b)\over
\sin(\lambda_a-\lambda_b-i\phi)}
\nonumber\\
&\times
\prod_{j\in \alpha^+}{\sin^{j-1}(\mu_j'+3i{\phi\over 2})
\sin^{m-j}(\mu_j'+i{\phi\over 2})\over \vartheta_1^m(\mu_j'+i\phi/2)}
\prod_{j\in \alpha^-}{\sin^{j-1}(\mu_j-i{\phi\over 2})
\sin^{m-j}(\mu_j+i{\phi\over 2})\over \vartheta_1^m(\mu_j+i\phi/2)},
\end{align}
where
\begin{align}
&\Delta=\cosh\phi,\quad q=e^{-\phi},\quad C_m=\prod^\infty_{n=1}\Big({1-q^{2n}\over 1+q^{2n}}\Big)^2\Big[2q^{1/4}
\prod^\infty_{n=1}(1-q^{2n})^3\Big]^{\frac{m(m+1)}{2}-1}=
\frac{(\vartheta'_1(0))^{\frac{m(m+1)}{2}}}{\vartheta_2(0)},\nn\\
&\vartheta_1(x)=q^{1/4}
\sum_{n=0}^\infty (-1)^nq^{n(n+1)/2}\sin(2n+1)x, \quad
\vartheta_2(x)=q^{1/4}\sum_{n=0}^\infty q^{n(n+1)/2}\cos(2n+1)x.
\end{align}
Then these equations are written as follows
\begin{align}
&F\Bigl({
\begin{matrix}
{\epsilon_1,\epsilon_2,...,\epsilon_m}\cr {\epsilon_1',\epsilon_2',...,\epsilon_m'}
\end{matrix}}
\Bigl)=\prod^m_{j=1}\int^{\pi/2}_{-\pi/2}\frac{{\rm d}\lambda_j}{2\pi}
U_m(\lambda_1,...,\lambda_m){\prod_{j=1}^m\sin^{y_j-1}(\lambda_j+i{\phi\over 2})
\sin^{m-y_j}(\lambda_j-i{\phi\over 2})
\over\prod_{a>b}
\sin(\lambda_a-\lambda_b-i f_{ab}\phi )},\label{mvP}\\
&U_m(\lambda_1,...,\lambda_m)=
i^mC_m
{\prod_{a>b} \vartheta_1(\lambda_a-\lambda_b)\over
\prod^m_{j=1}\vartheta_1^m(\lambda_j+i\phi/2)}
\vartheta_2(\sum_{j=1}^m(\lambda_j+i\phi/2)),
\end{align}
where $f_{ab}$ and $y_j$ are given by (\ref{fyj}).
In Appendix \ref{A1} we show that 
\begin{align}
{ U}_m(\lambda_1,...,\lambda_m)={1\over \prod_{r=1}^{m-1} r!}
\prod_{j>k}({\partial\over\partial \lambda_j}-{\partial\over\partial \lambda_k})
{U}_1(\lambda_1){U}_1(\lambda_2)...{U}_1(\lambda_m).\label{Um1}
\end{align}

If we put $\lambda_j\to x_j\phi$ this is written as 
\begin{align}
&\prod^m_{j=1}\int^{Q/2}_{-Q/2}{\rm d}x_j{\bf W}_m(x_1,...,x_m)
{\displaystyle\prod_{j=1}^m\sin^{y_j-1}\phi(x_j+i{1\over 2})\sin^{m-y_j}\phi(x_j-i{1\over 2})
\over\phi^{m(m-1)/2}\prod_{a>b}\sin\phi(x_a-x_b-i f_{ab})},\\
&Q=\pi/\phi, \quad{\bf W}_m(x_1,...,x_m)=\phi^{m(m-1)/2}U_m(\phi x_1,...,\phi x_m).
\end{align}
We investigate the properties of ${\bf W}_1$,
\begin{align}
{\bf W}_1(x)=i{\vartheta'_1(0)\vartheta_2(\phi(x+i/2))\over
\vartheta_2(0)\vartheta_1(\phi(x+i/2))}={K_l\over \pi Q}{\rm dn}({2K_l x\over Q},l),
\end{align}
where ${\rm dn}(z,l)$ is Jacobian elliptic function with
$${K(\sqrt{1-l^2})\over K(l)}={\phi\over\pi},\quad K(l)=\int^{\pi/2}_0{1\over\sqrt{1-l^2\sin^2 k}}.$$
This function has poles at $x=kQ+i(l-\frac{1}{2})$ with residue
$(-1)^li$ and zeros at $\lambda=(k-\frac{1}{2})Q+i(l+\frac{1}{2})$.  
On the other hand $\sum_{n=-\infty}^\infty \frac{1}{2}{\rm sech}(\pi(x+nQ))$
has the same poles and zeroes. Then we have
\begin{align}
{\bf W}_1(x)=\sum_{n=-\infty}^\infty W_1(x+nQ).
\label{U1}\end{align}
Using (\ref{Um1}) we have
\begin{align}
{\bf W}_m(x_1,...,x_m)=D_m{\bf W}_1(x_1){\bf W}_1(x_2)...{\bf W}_1(x_m).
\label{Um}\end{align}
Then we have
\begin{align}
&{\bf W}_m(x_1,...,x_m)=\sum_{n_1,...,n_m} W_m(x_1+n_1Q,x_2+n_2Q,...,x_m+n_mQ).
\end{align}
The weight function in massless case is trigonometric. But in massive case it becomes elliptic 
and satisfies the following properties,
\begin{align}
& {\bf W} _m(x_1,x_2,...,x_j,...,x_k,...,x_m)=-{\bf W} _m(x_1,x_2,...,x_k,...,x_j,...,x_m), \label{ex}\\
&{\bf W} _m(x_1,...,x_j,...,x_m)={\bf W} _m(x_1,...,x_j+Q,...,x_m)=-{\bf W} _m(x_1,...,x_j+i,...,x_m),\label{peri}\\
&{\bf W} _m(x_1,...,x_j,...,x_m)=(-1)^{m(m-1)/2}{\bf W} _m(-x_1,...,-x_j,...,-x_m).\label{rev}
\end{align}
We should note that weight function is doubly periodic but the integrand is not doubly periodic. 
The correlation function has the following properties
\begin{align}
&F\Bigl({
\begin{matrix}
{\epsilon_1,\epsilon_2,...,\epsilon_m}\cr {\epsilon_1',\epsilon_2',...,\epsilon_m'}
\end{matrix}}
\Bigr)=F\Bigl({
\begin{matrix}
{-\epsilon_1,-\epsilon_2,...,-\epsilon_m}\cr {-\epsilon_1',-\epsilon_2',...,-\epsilon_m'}
\end{matrix}}
\Bigr)=F\Bigl({
\begin{matrix}
{\epsilon_1',\epsilon_2',...,\epsilon_m'}\cr {\epsilon_1,\epsilon_2,...,\epsilon_m}
\end{matrix}}\Bigr)=F\Bigl({
\begin{matrix}
{\epsilon_m,\epsilon_{m-1},...,\epsilon_1}\cr {\epsilon_m',\epsilon_{m-1}',...,\epsilon_1'}
\end{matrix}}\Bigr),\\
&F\Bigl({
\begin{matrix}
{\epsilon_1,\epsilon_2,...,\epsilon_{m-1},+}\cr {\epsilon_1',\epsilon_2',...,\epsilon_{m-1}',+}
\end{matrix}}\Bigr)+F\Bigl({
\begin{matrix}
{\epsilon_1,\epsilon_2,...,\epsilon_{m-1},-}\cr {\epsilon_1',\epsilon_2',...,\epsilon_{m-1}',-}
\end{matrix}}\Bigr)=F\Bigl({
\begin{matrix}
{\epsilon_1,\epsilon_2,...,\epsilon_{m-1}}\cr {\epsilon_1',\epsilon_2',...,\epsilon_{m-1}'}
\end{matrix}}\Bigr).
\end{align}

\section{Reduction to one dimensional integral at $m=2$ and $3$}
At first we consider EFP defined by (\ref{Pm}). This is equal to 
$F({\begin{matrix}{+,+,...,+}\cr {+,+,...,+}\end{matrix}})$.
As $y_j=m+1-j, f_{ab}=1$ and we
have the integral representation 
\begin{align}
P(m)= \int_{-Q/2}^{Q/2}{\rm d}^m x 
{\bf W}_m
\frac{\displaystyle\prod_{k=1}^{m} h^{m-k}\left(x_k + {\rm i}/2 \right)h^{k-1}\left(x_k - {\rm i}/2 \right)}
{\displaystyle\prod_{a>b} h\left(x_a-x_b - {\rm i} \right)},\quad {\rm with}\quad h(x)\equiv \sin(\phi x)/\phi. 
\label{EFPintegral}
\end{align}
Recently there have been an increasing number of researches concerning the properties 
of the EFP.
Particularly, in the isotropic limit ${\Delta \to 1 (\phi \to 0)}$, the general 
method to evaluate the multiple integral was recently developed by Boos et al\cite{Boos01,Boos02,BKNS02}. 

It is important to note that ${P(2)}$ and ${P(3)}$ are related to the 
nearest and next nearest-neighbor correlation functions :
\begin{align}
\langle S_{j}^z S_{j+1}^z \rangle &= P(2)-1/4, \label{P2} \\ 
\langle S_{j}^z S_{j+2}^z \rangle &= 2 (P(3)-P(2)+1/8). 
\label{P3} 
\end{align}
We also consider transverse correlation 
functions ${\langle S_{j}^z S_{j+1}^z\rangle}$ and
${\langle S_{j}^z S_{j+2}^z \rangle}$.  Then we should evaluate following 
four integrals 
\begin{align}
&P(2) =   \int_{-Q/2}^{Q/2}  
{\rm d}^2 x
{\bf W}_2
\frac{h( x_1 + {\rm i}/2 ) h(x_2-{\rm i}/2 )}{h(x_2-x_1 -i)}, \label{P2a}\\
&\langle S_{j}^x S_{j+1}^x \rangle = \frac{1}{2}  \int_{-Q/2}^{Q/2}  
{\rm d}^2 x
{\bf W}_2
\frac{h( x_1 + {\rm i}/2 ) h(x_2-{\rm i}/2 )}{h(x_2-x_1 )}, 
\label{transverse1}\\
&P(3) = 
\int_{-Q/2}^{Q/2} {\rm d}^3 x{\bf W}_3
\frac{h^{2} (x_1 + \frac{\rm i}{2})
h(x_2 + \frac{\rm i}{2} ) h(x_2 - \frac{\rm i}{2} ) 
h^2(x_3 - \frac{\rm i}{2} )}{h(x_2-x_1- {\rm i}) h(x_3-x_1- {\rm i} ) 
h(x_3-x_2 - {\rm i})},
\label{P3a}\\
&\langle S_{j}^x S_{j+2}^x \rangle =\displaystyle 
\int_{-Q/2}^{Q/2} {\rm d}^3 x{\bf W}_3
\frac{h^{2} (x_1 + \frac{\rm i}{2})
h(x_2 + \frac{\rm i}{2} ) h(x_2 - \frac{\rm i}{2} ) 
h^2(x_3 - \frac{\rm i}{2} )}{h(x_2-x_1 ) h(x_3-x_1 ) 
h(x_3-x_2 - {\rm i})}.
\label{transverse2}
\end{align}
In Appendix \ref{A2} we transform these to the canonical forms. The  
two-dimensional integrals of (\ref{P2a}) and (\ref{transverse1}) can be reduced to 
one-dimensional ones,
\begin{align}
&P(2)= \frac{1}{2}+ \int_{- \infty+ {\rm i}/2}^{\infty+ {\rm i}/2} 
\frac{{\rm d} x}{\sinh \pi x}{1\over 2} (-{x\over \sin^2 \phi x} +\cot(\phi x)\coth\phi ) ,\label{P2c}\\
&\langle S_{j}^x S_{j+1}^x \rangle = \int_{- \infty+ {\rm i}/2}^{\infty+ {\rm i}/2} 
\frac{{\rm d} x}{\sinh \pi x}{1\over 4}({x\over \sin^2 \phi x} \cosh\phi
-{\cot(\phi x)\over \sinh\phi}). \label{transverse1c}
\end{align}
These coincide with known results. The main results are that 
{\it three-dimensional integrals for ${P(3)}$ and 
${\langle S_{j}^x S_{j+2}^x \rangle}$ can also be reduced 
to one-dimensional integrals}. In other words, we have succeeded in performing 
the integrals for ${P(3)}$ and ${\langle S_{j}^x S_{j+2}^x \rangle}$:
\begin{align}
&P(3)= \frac{1}{2} 
+ \int_{- \infty+ {\rm i}/2}^{\infty+ {\rm i}/2} 
\frac{{\rm d} x}{\sinh \pi x}{1\over 8} \Bigg[ 
\frac{x}{\sin^2(\phi x)}\Big({3\sinh^2\phi\over\sin^{2}(\phi x)}-5-\cosh 2\phi\Big)
-\cot(\phi x)\Big({3\tanh\phi\over\sin^{2}(\phi x)}-{4+8\cosh 2\phi\over \sinh 2\phi}\Big) 
\Bigg], \label{P3new}\\
&\langle S_{j}^x S_{j+2}^x \rangle = \int_{- \infty+ {\rm i}/2}^{\infty+ {\rm i}/2} 
\frac{{\rm d} x}{\sinh \pi x}{1\over 8} \Bigg[ -\frac{x}{\sin^2(\phi x)}
\Big(\frac{3\sinh^2\phi}{\sin^{2}(\phi x)}+1-3\cosh 2\phi\Big)
+\cot(\phi x)\Big({3\cosh(2\phi) \tanh(\phi)\over\sin^{2}(\phi x) }
-{4\over\sinh^{}(2\phi)}\Big)
\Bigg]. \label{transnew}
\end{align}
Using equations (\ref{P2}, \ref{P3}) we have
\beqa
&&\langle S_{j}^z S_{j+1}^z \rangle =\frac{1}{4}+ \int_{- \infty+ {\rm i}/2}^{\infty+ {\rm i}/2} 
\frac{{\rm d} x}{\sinh \pi x}{1\over 2} (\cot(\phi x)\coth(\phi)-{x\over \sin^2 \phi x}) ,\label{sz1}\\
&&\langle S_{j}^z S_{j+2}^z \rangle=  \frac{1}{4} 
+ \int_{- \infty+ {\rm i}/2}^{\infty+ {\rm i}/2} 
\frac{{\rm d} x}{\sinh \pi x}{1\over 4} \Bigg[ 
\frac{x}{\sin^2(\phi x)}\Big({3\sinh^2\phi\over\sin^{2}(\phi x)}-1-\cosh 2\phi
\Big)
-\cot(\phi x)\Big({3\tanh\phi\over\sin^{2}(\phi x)}-4\coth2\phi
\Big) 
\Bigg].\label{sz2}
\enqa
\begin{table} 
\caption{Estimated values of correlation function $  \bra S^z_0 S^z_2\ket$ 
from finite rings and our calculation }
{
\begin{tabular}{llllllll}
$N$           &{$\Delta=1.0$}&{$\Delta=1.5$}& {$\Delta=2.0$} &{$\Delta=2.5$} &{$\Delta=3.0$}  &{$\Delta=3.5$} & {$\Delta=4.0$} \\
\hline
20  &         0.06135072260 & 0.1031381904 & 0.1473943448 & 0.1795178252 &  0.1993399825 &  0.2119381546 & 0.2204021144 \\ 
22  &         0.06123184969 & 0.1033007174 & 0.1480705446 & 0.1799480453 &  0.1995078997 &  0.2119982680 & 0.2204241750 \\
24  &         0.06114201704 & 0.1034578190 & 0.1486278313 & 0.1802452204 &  0.1996070522 &  0.2120292690 & 0.2204342780 \\
26  &         0.06107247281 & 0.1036073511 & 0.1490859151 & 0.1804501423 &  0.1996656632 &  0.2120452914 & 0.2204389106 \\
28  &         0.06101753236 & 0.1037485416 & 0.1494616068 & 0.1805913409 &  0.1997003229 &  0.2120535929 & 0.2204410300 \\
30  &         0.06097337258 & 0.1038813030 & 0.1497692404 & 0.1806886032 &  0.1997209063 &  0.2120579113 & 0.2204420230 \\
extra.&       0.06067894953 & 0.1061687241 & 0.1508088848 & 0.1809161322 &  0.1999716192 &  0.2120736609 & 0.2204679674 \\
exact &       0.06067976996 & 0.1061446753 & 0.1511283748 & 0.1809054601 &  0.1997511670 &  0.2120626058 & 0.2204428632 \\
\hline
\end{tabular}\label{tab1}
}
\caption{Estimated values of correlation function $  \bra S^x_0 S^x_2\ket$ 
from finite rings and our calculation }
{
\begin{tabular}{lllllll}
$N$           &{$\Delta=1.5$}& {$\Delta=2.0$} &{$\Delta=2.5$} &{$\Delta=3.0$}  &{$\Delta=3.5$} & {$\Delta=4.0$} \\
\hline
20  &          0.04336844424 & 0.02832886249  & 0.01874771832 &  0.01321049421 &  0.00981186886 & 0.00757459251 \\ 
22  &          0.04318176477 & 0.02806630512  & 0.01860734227 &  0.01315890868 &  0.00979394498 & 0.00756813199 \\
24  &          0.04303078324 & 0.02785949276  & 0.01851259253 &  0.01312895682 &  0.00978483286 & 0.00756521107 \\
26  &          0.04290579437 & 0.02769531699  & 0.01844841522 &  0.01311148673 &  0.00978017739 & 0.00756388405 \\
28  &          0.04280033153 & 0.02756430324  & 0.01840481529 &  0.01310126617 &  0.00977778854 & 0.00756328233 \\
30  &          0.04270993496 & 0.02745937089  & 0.01837513880 &  0.01309525214 &  0.00977655328 & 0.00756300336 \\
extra.&        0.04170186316 & 0.02715342577  & 0.01832502644 &  0.01302827108 &  0.00976919713 & 0.00755456686 \\
exact &        0.04171854326 & 0.02702161935  & 0.01831051051 &  0.01308654379 &  0.00977523196 & 0.00756276973 \\
\hline
\end{tabular}\label{tab2}
}
\end{table}
}\end{widetext}

\begin{figure}[hbtp]
\includegraphics[width=8.5cm]{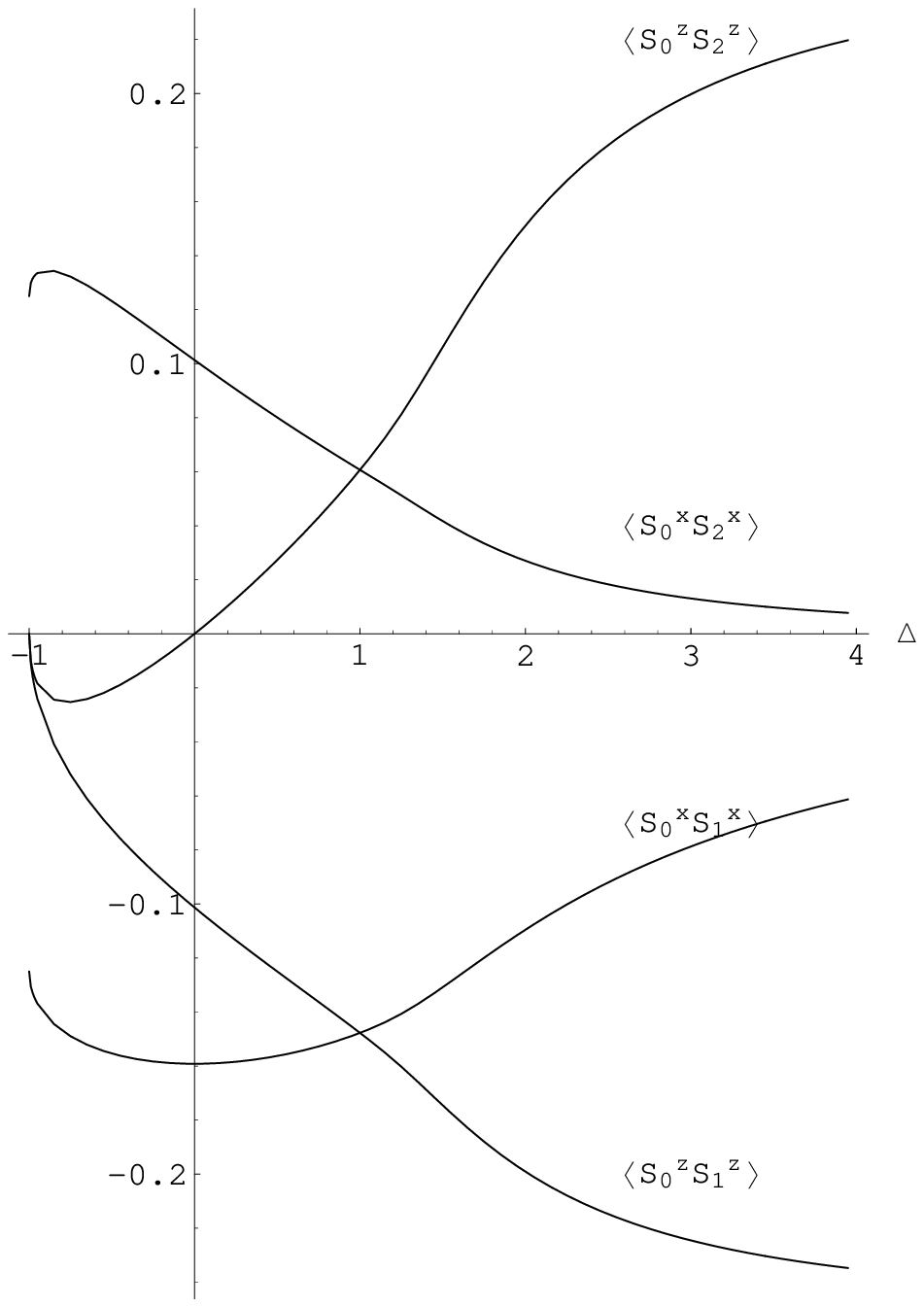}
\caption{Nearest-neighbor and next nearest neighbor correlation functions for the XXZ chain. 
We calculated $\bra S_j^zS_{j+1}^z\ket$, $\bra S_j^xS_{j+1}^x\ket$, $\bra S_j^zS_{j+2}^z\ket$ and $\bra S_j^xS_{j+2}^x\ket$ }
\label{fig:fig1}
\end{figure}

\section{Numerical results and conclusion}
We calculate numerically the second neighbor correlations $  \bra S^z_0 S^z_2\ket$ 
and $  \bra S^x_0 S^x_2\ket$ for finite rings with $N=20,22,24,26,28,30$. In Table (\ref{tab1}) 
and (\ref{tab2}) results of (\ref{sz2}) and (\ref{transnew}) are compared with numerical results. 
They coincide very well.

In the previous paper we treated the second neighbor correlations for massless region $1>\Delta>-1$. 
In this paper we have shown that almost the same method can be applied for massive region ${\Delta>1}$.
This is the generalization of $\Delta=1$ result 
\beq
\langle S_j^{z} S_{j+2}^z\rangle=\frac{1}{12}-\frac{4}{3}\ln2+\frac{3}{4}\zeta(3).
\enq
The multiple integrals for the correlation 
functions of the ${XXZ}$ chain at the massive region 
can be reduced to the one-dimensional ones in the case of the next 
nearest-neighbor correlation functions. This property will be generalized 
to the third neighbor or longer distance correlations. 
In our future work, we are particularly interested in calculating the 
third-neighbor correlation functions ${\langle S_{j}^x S_{j+3}^x \rangle}$ 
and ${\langle S_{j}^z S_{j+3}^z \rangle}$ for general ${\Delta}$. 
 For ${\Delta=1}$, they were recently calculated in \cite{Sakai03} 
\beqa
&\langle  S^z_j S^z_{j+3}\rangle=\frac{1}{12}-3\ln 2+\frac{37}{6}\zeta(3)
-\frac{14}{3}\zeta(3)\ln 2\nn\\
&-\frac{3}{2}\zeta(3)^2 -\frac{125}{24}\zeta(5)
+\frac{25}{3}\zeta(5)\ln 2.
\enqa
from the multiple integrals. This should be generalized to arbitrary $\Delta$. 
\begin{acknowledgments}
We are grateful to K. Sakai for valuable discussions. 
The numerical data for finite rings have been obtained 
by the Fortran package TITPACK Ver. 2.
This work is in part 
supported by Grant-in-Aid for the Scientific Research (B) No.~14340099 
from the Ministry of Education, Culture, Sports, Science and Technology,  Japan. 
One of the authors (GK) is supported by the JSPS
research fellowships for young scientists. MS is supported by Grant-in Aid for Young Scientists (B) 
No. 14740228.

\end{acknowledgments}
\appendix

\section{Antisymmetric differential operator}\label{A1}
Elliptic theta function satisfies the following quasi double periodicity
\begin{align}
\vartheta_1(\lambda+\pi)=-\vartheta_1(\lambda),\quad \vartheta_1(\lambda+i\phi)
=-e^\phi e^{-2i\lambda}\vartheta_1(\lambda).\label{qperi}
\end{align}
We consider 
$$f(\lambda_1,\lambda_2)=({\partial\over\partial\lambda_2}-{\partial\over\partial\lambda_1})
U_1(\lambda_1)U_1(\lambda_2).$$
Using (\ref{qperi}) and $\vartheta_2(\lambda)=\vartheta_1(\lambda+\pi/2)$ 
we can show that $U_1(\lambda)$ and $f(\lambda_1,\lambda_2)$ are elliptic 
functions which satisfy
\begin{align}
&U_1(\lambda_2)=-U_1(\lambda_2+i\phi)=U_1(\lambda_2+\pi),\nonumber\\&
f(\lambda_1,\lambda_2)=-f(\lambda_1,\lambda_2+i\phi)=f(\lambda_1,\lambda_2+\pi).\label{peri2}
\end{align}
As $f(\lambda_1,\lambda_2)=-f(\lambda_2,\lambda_1)$, $f$ has zeros at 
$\lambda_1+k\pi+il\phi$ as function of $\lambda_2$. It has poles 
at $k\pi+i(l-1/2)\phi$ with order 2.  Then function $f$ should be factorized as follows
\begin{align}
f(\lambda_2)=C(\lambda_1){\vartheta_1(\lambda_2-\lambda_1)\vartheta_1(\lambda_2-\alpha)
\over \vartheta_1(\lambda_2+i\phi/2)^2}.
\end{align}
Here $\alpha$ is another zero. Using (\ref{qperi}) we have 
$f(\lambda_2+i\phi)=\exp(2i(\alpha+\lambda_1+i\phi))f(\lambda_2)$. 
By the condition (\ref{peri2}) $\alpha$  must be $\pi/2-\lambda_1-i\phi$.
The residue of $U_1(\lambda)$ at $\lambda=-i\phi/2$ is $i$. Then we have
$$\lim_{\lambda_2\to -i\phi/2}(\lambda_2+i\frac{\phi}{2})^2 f(\lambda_1,\lambda_2)
=-iU_1(\lambda_1).$$
By this condition we find
\begin{widetext}
\begin{align}
C(\lambda_1)=iU_1(\lambda_1){(\vartheta_1'(0))^2\over \vartheta_1(\lambda_1+i\phi/2)
\vartheta_1(\lambda_1+i\phi/2+\pi/2)},
\end{align}
and therefore $f(\lambda_1,\lambda_2)=U_2(\lambda_1,\lambda_2)$. Thus we 
have proven (\ref{Um}) for $m=2$.\par
To prove for $m\ge 3$ we assume that it stands for $U_{m-1}$. Define
\begin{align}
f_m(\lambda_1,...,\lambda_m)={1\over (m-1)!}\prod_{l=1}^{m-1}({\partial\over\partial\lambda_m}
-{\partial\over\partial\lambda_l})
U_1(\lambda_m)U_{m-1}(\lambda_1,...,\lambda_{m-1}).
\end{align}
This function has the following properties:
\begin{align}
f_m(\lambda_1,...,\lambda_m)=-f(\lambda_1,...,\lambda_m+i\phi)
=f(\lambda_1,...,\lambda_m+\pi).\label{perim}
\end{align}
As $f_m$ is antisymmetric for exchange of variables we know $m-1$ zeroes and $m$ fold 
pole:
\beq
f_m(\lambda_m)=C(\lambda_1,...,\lambda_{m-1}){\displaystyle
\prod_{l=1}^{m-1}\vartheta_1(\lambda_m-\lambda_l)
\vartheta_1(\lambda_m-\alpha)
\over \vartheta_1(\lambda_m+i\phi/2)^m}.
\enq
$\alpha$ must be $\pi/2-\sum_{l=1}^{m-1}\lambda_l-im\phi/2$ by the condition (\ref{perim}).
Using 
$$\lim_{\lambda_m\to -i\phi/2}(\lambda_m+i\frac{\phi}{2})^m f_m(\lambda_1,...,\lambda_m)
=(-1)^{m-1}iU_{m-1}(\lambda_1,...,\lambda_{m-1}),$$
we can determine $C$,
\begin{align}
C(\lambda_1,...,\lambda_{m-1})=iU_{m-1}(\lambda_1,...,\lambda_{m-1})
{(\vartheta_1'(0))^m\over \vartheta_1(\displaystyle\sum_{l=1}^{m-1}(\lambda_l+i\phi/2)+\pi/2)
\prod_{l=1}^{m-1}\vartheta_1(\lambda_l+i\phi/2)}.
\end{align}
Thus we show 
\begin{align}
{ U}_m(\lambda_1,...,\lambda_m)={1\over \prod_{r=1}^{m-1} r!}
\prod_{j>k}({\partial\over\partial \lambda_j}-{\partial\over\partial \lambda_k})
{U}_1(\lambda_1){U}_1(\lambda_2)...{U}_1(\lambda_m).
\end{align}
for general $m$ stands if it stands for $m-1$. Putting $\lambda_j=\phi x_j$ we obtain (\ref{Um}).

\section{Some integral formula}
For the integration of canonical forms following formula is important,
\begin{align}
&\int^{Q/2}_{-Q/2}{\rm d}x_1  e^{2in\phi x_1}{\bf W}_1(x_1+\epsilon_1)
=\int^{Q/2}_{-Q/2}{\rm d}x_1  e^{2in\phi x_1-\epsilon_1}{\bf W}_1(x_1)\nn\\
&=\int^{\infty}_{-\infty}{\rm d}x_1  e^{2in\phi( x_1-\epsilon_1)}{1\over 2}{\rm sech}(\pi x_1)
={\exp(-2in\phi\epsilon_1)\over 2\cosh n\phi}.\label{B1}
\end{align}
Here we have used (\ref{U1}). Next we consider the integral of $e^{2in\phi x_1}/( e^{2i\phi x_2}
-e^{2i\phi x_1})$:
\begin{align}
&\int^{Q/2}_{-Q/2}{\rm d} x_1{\rm d} x_2 {e^{2in\phi x_1}\over e^{2i\phi x_2}-e^{2i\phi x_1}}
{\bf W}_1(x_1+\epsilon_1){\bf W}_1(x_2+\epsilon_2)\nn\\
&=-i \int^{Q/2}_{-Q/2} \int^{Q/2}_{-Q/2}{\exp(i\phi((2n-1)x_1-x_2))\over 2\sin\phi(x_2-x_1)}
{\bf W}_1(x_1+\epsilon_1){\bf W}_1(x_2+\epsilon_2){\rm d} x_1{\rm d} x_2\nn\\
&=-i e^{i\phi(-(2n-1)\epsilon_1+\epsilon_2)}
\int^{Q/2}_{-Q/2} \int^{Q/2}_{-Q/2}
{ \exp(i\phi((2n-1)x_1-x_2))\over 2\sin\phi(x_2-x_1+\epsilon_1-\epsilon_2)}
{\bf W}_1(x_1){\bf W}_1(x_2){\rm d} x_1{\rm d} x_2 \nn
\end{align}
Using (\ref{U1}) we extend the region of integration:
\begin{align}
=-i e^{i\phi(-(2n-1)\epsilon_1+\epsilon_2)}
\int^{\infty}_{-\infty}\int^{\infty}_{-\infty} 
{\exp(i\phi((2n-1)x_1-x_2))\over 8\sin\phi(x_2-x_1+\epsilon_1-\epsilon_2)}
{\rm sech}\pi x_1 {\rm sech} \pi x_2 {\rm d} x_1{\rm d} x_2. \nn
\end{align}
The term ${\rm sech}\pi x_1{\rm sech}\pi x_2$ can be written as $(\tanh\pi x_2-\tanh\pi x_1)/\sinh\pi(x_2-x_1)$. 
Putting $y_1=x_1+x_2, y_2=x_2-x_1$ we have
\begin{align}
=-i e^{i\phi(-(2n-1)\epsilon_1+\epsilon_2)}\int^\infty_{-\infty}
{{\rm d}y_2 \exp(-i\phi n y_2))\over 16\sinh\pi y_2 
\sin\phi(y_2+\epsilon_1-\epsilon_2)}
\int^\infty_{-\infty} e^{i(n-1)\phi y_1} (\tanh{\pi\over 2}(y_1+y_2)-\tanh{\pi\over 2}(y_1-y_2))
{\rm d} y_1\nn.
\end{align}
For $n\ne 1$ the last integral is estimated as follows:
\begin{align}
&\int^\infty_{-\infty} e^{i(n-1)\phi y_1} (\tanh{\pi\over 2}(y_1+y_2)-\tanh{\pi\over 2}(y_1-y_2))
{\rm d} y_1\nn\\
&=-\frac{\pi}{2}\int^\infty_{-\infty}\frac{e^{i(n-1)\phi y_1}-1}{(n-1)\phi i}
\Big(\frac{1}{\cosh ^2 {\pi\over 2}(y_1+y_2)}%
- \frac{1}{\cosh ^2 {\pi\over 2}(y_1-y_2)}\Big){\rm d} y_1
\nn\\
&=\frac{\pi}{2}\int^\infty_{-\infty}{e^{i(n-1)\phi(y_1+y_2)}-e^{i(n-1)\phi(y_1-y_2)}\over (n-1)\phi i}
{{\rm d} y_1\over \cosh ^2 {\pi\over 2}y_1}=\pi{\sin(n-1)\phi y_2\over (n-1)\phi}\int^\infty_{-\infty} e^{i\phi(n-1)y_1}
{{\rm d} y_1\over \cosh ^2 {\pi\over 2}y_1}\nn\\
&={4\sin(n-1)\phi y_2\over \sinh (n-1)\phi}.\nn
\end{align}
For $n=1$ this is $4y_2$. Then only 
integral with respect to $y_2$ remains. To circumvent the poles on the real axis we change the path of 
integral from $[-\infty,\infty]$ to $[-\infty+i/2,\infty+i/2]$. Thus we have
\begin{align}
&\int^{Q/2}_{-Q/2}{\rm d} x_1{\rm d} x_2 {e^{2in\phi x_1}\over e^{2i\phi x_2}-e^{2i\phi x_1}}
{\bf W}_1(x_1+\epsilon_1){\bf W}_1(x_2+\epsilon_2)\nn\\
&=\left\{\begin{array}{@{\,}ll}&
-i e^{i\phi(-(2n-1)\epsilon_1+\epsilon_2)}{\displaystyle
\int^{\infty+i/2}_{-\infty+i/2}
{{\rm d}x \exp(-ni\phi x)\over 4\sinh\pi x \sin\phi(x+\epsilon_1-\epsilon_2)} 
{\sin(n-1)\phi x\over \sinh (n-1)\phi}},\quad {\rm for}\quad n\ne 1,\\\\
&
-i e^{i\phi(-\epsilon_1+\epsilon_2)}{\displaystyle
\int^{\infty+i/2}_{-\infty+i/2}
{{\rm d}x  \exp(-i\phi x)\over 4\sinh\pi x \sin\phi(x+\epsilon_1-\epsilon_2)}x,}
\quad {\rm for}\quad n=1.\end{array}\right.\label{B2}
\end{align}
If the integrand is given by a canonical form we can estimate the integral by these formulae.

\section{Canonical form of integrand}\label{A2}
The weight function ${\bf W}_m$ is elliptic function with periodicity $Q,2i$ and the integrand is 
periodic function with periodicity $Q$. Then integrand is expressed by $z_l\equiv\exp(2i\phi x_l+\phi)$.
If integrand has the form $(z_1-1)^mf(z_1,z_2,...,z_m)$ and $f$ has no singularity as a function of $z_1$, 
the $m$-th pole of  ${\bf W}_m$ at $x_1=i/2$ is cancelled and shift of path from $[-Q/2,Q/2]$ to $[-Q/2+i,Q/2+i]$
becomes possible:
\begin{align}
\int^{Q/2}_{-Q/2}{\rm d}x_1{\bf W}_m(z_1-1)^mf(z_1,z_2,...)
=-\int^{Q/2}_{-Q/2}{\rm d}x_1{\bf W}_m(q^2z_1-1)^mf(q^2z_1,z_2,...).
\end{align}
Here we have used (\ref{peri}). In the same way we have
\begin{align}
\int^{Q/2}_{-Q/2}{\rm d}x_1{\bf W}_m(q^2z_1-1)^mf(z_1,z_2,...)
=-\int^{Q/2}_{-Q/2}{\rm d}x_1{\bf W}_m(z_1-1)^mf(q^{-2}z_1,z_2,...).
\end{align}
Actually $f$ might have single pole at $z_1=q^{2n}z_2,\quad n={\rm integer}$, because 
${\bf W}_m$ has single zeros at these points. Then we have
\begin{align}
&\int^{Q/2}_{-Q/2}{\rm d}x_1{\bf W}_m{(z_1-1)^m\over z_1-q^2z_2}f(z_1,z_2,...)
=-\int^{Q/2}_{-Q/2}{\rm d}x_1{\bf W}_m{(q^2z_1-1)^m\over q^2(z_1-z_2)}f(q^2z_1,z_2,...),\\
&\int^{Q/2}_{-Q/2}{\rm d}x_1{\bf W}_m{(q^2z_1-1)^m\over q^2z_1-z_2}f(z_1,z_2,...)
=-\int^{Q/2}_{-Q/2}{\rm d}x_1{\bf W}_m{(z_1-1)^m\over z_1-z_2}f(q^{-2}z_1,z_2,...).\end{align}
From (\ref{ex}) we have
\begin{align}
\int^{Q/2}_{-Q/2}{\rm d}^m x{\bf W}_m g(z_1,z_2,z_3,...)=-\int^{Q/2}_{-Q/2}{\rm d}^m x{\bf W}_m g(z_2,z_1,z_3,...).
\end{align}
Suffix 1 and 2 can be replaced by general $j\ne k$ at $1\le j,k\le m$. 
These rules are sufficient to simplify the integrand. Hereafter we write $A\sim B$ if 
$$\int^{Q/2}_{-Q/2}{\rm d}^m x{\bf W}_m A=\int^{Q/2}_{-Q/2}{\rm d}^m x{\bf W}_m B.$$

At first we consider two-dimensional integral. Equation (\ref{P2a}) and 
(\ref{transverse1}) are rewritten in terms of $q\equiv \exp(-\phi)$ and
$z_l\equiv\exp(2i\phi x_l+\phi)$,
\begin{align}
P(2) =   \int_{-Q/2}^{Q/2}  
{\rm d} x_1{\rm d} x_2 {\bf W}_2
\frac{(q^2 z_1-1)(z_2-1)}{2i\phi(z_2-q^2 z_1)}. \label{P2b}
\end{align}
The integrand is transformed as follows
\begin{align}
&\frac{(q^2 z_1-1)(z_2-1)}{2i\phi(z_2-q^2 z_1)}={1\over 2i\phi}(-z_2+1+{(z_2-1)^2\over z_2-q^2 z_1})
\sim{1\over 2i\phi}(-z_2-{(q^2 z_2-1)^2\over q^2 (z_2-z_1)}).\label{p2k}
\end{align}
We can show that $z_1^2/(z_2-z_1)\sim -z_1+q^{-2}/(z_2-z_1)$, because
\begin{align}
&{(z_1-1)^2\over z_2-z_1}\sim -{(q^2z_1-1)^2\over z_2-q^2z_1}=q^2z_1+z_2-2-{(z_2-1)^2\over z_2-q^2z_1}\nn\\
&\sim-(1-q^2)z_1+{(q^2z_2-1)^2\over q^2(z_2-z_1)}\sim-(1-q^2)z_1+{(q^2z_1-1)^2\over q^2(z_2-z_1)}.
\end{align}
Then we have the canonical form
\begin{align}
{\rm (\ref{p2k})}\sim{1\over 2i\phi}((1+q^2)z_1+{2z_1-1-q^{-2}\over z_2-z_1}).\label{P2e}
\end{align}
Using (\ref{B1}) and (\ref{B2}) we evaluate
$\int^{Q/2}_{-Q/2}{\rm d} x_1{\rm d} x_2 {\bf W}_1(x_1+\epsilon_1){\bf W}_1(x_2+\epsilon_2){\rm(\ref{P2e})}$:
\begin{align}
{e^{-2i\phi \epsilon_1}\over 4i\phi}+{1\over 4\phi} \int^{\infty+i/2}_{-\infty+i/2}
{{\rm d}x\over\sinh \pi x\sin\phi(x+\epsilon_1-\epsilon_2)}\Big[-x e^{-i\phi(x+\epsilon_1-\epsilon_2)}
+e^{i\phi(\epsilon_1+\epsilon_2)}\coth\phi\sin\phi x\Big].\label{P2d}
\end{align}
Using (\ref{Um}) we have
\begin{align}
&P(2)=\lim_{\epsilon_1,\epsilon_2\to 0}(\partial_2-\partial_1)({\rm \ref{P2d}}) =(\ref{P2c}).
\end{align}

(\ref{transverse1}) is rewritten 
\begin{align}
\langle S_{j}^x S_{j+1}^x \rangle =  \int_{-Q/2}^{Q/2}  
{\rm d} x_1{\rm d} x_2 {\bf W}_2
\frac{(q^2 z_1-1)(z_2-1)}{4i\phi q(z_2-z_1)}.
\label{transverse1a}
\end{align}
The canonical form of integrand is
\begin{align}
\frac{1}{4i\phi q}{-(1+q^2)z_1+2\over z_2-z_1}.
\end{align}
Then we have
\begin{align}
&\langle S_{j}^x S_{j+1}^x \rangle =\lim_{\epsilon_1,\epsilon_2\to 0}(\partial_2-\partial_1)
{1\over 8\phi} \int^{\infty+i/2}_{-\infty+i/2}
{{\rm d}x\over\sinh \pi x\sin\phi(x+\epsilon_1-\epsilon_2)}
\Big[x \cosh\phi e^{-i\phi(x+\epsilon_1-\epsilon_2)}
-e^{i\phi(\epsilon_1+\epsilon_2)}\frac{\sin\phi x}{\sinh\phi}\Big]=(\ref{transverse1c}).
\end{align}

Equation (\ref{P3a}) is
\begin{align}
P(3) &= {i\over 8\phi^3}
\int_{-Q/2}^{Q/2} {\rm d}^3 x{\bf W}_3
\frac{(q^2z_1-1)^2(q^2z_2-1)(z_2 -1) 
(z_3 -1)^2}{(z_2-q^2z_1)(z_3-q^2z_1) 
(z_3-q^2z_2)},
\label{P3b}
\end{align}
The integrand is devided as follows
\begin{align}
&{(q^2z_1-1)^2(q^2z_2-1)(z_2 -1)(z_3 -1)^2\over (z_2-q^2z_1)(z_3-q^2z_1)(z_3-q^2z_2)}=
{(q^2z_1-1)^2(q^2z_2-1)(z_2 -1)(z_3 -1)^2\over z_2(1-q^2)}\nn\\
&\times\Bigl( {1\over (z_3-q^2z_1)(z_3-q^2z_2)}+{1\over (z_3-q^2z_1)(z_2-q^2z_1)}
-{1\over (z_3-q^2z_2)(z_2-q^2z_1)}\Bigr)\nn
\end{align}
The first and second terms are symmetrized by  the exchange of variables and the third term is 
devided to four terms
\begin{align}
&\sim-(q^2z_1-1)(q^2z_2-1){(z_3-1)^2(1+q^2 (z_1 z_2-z_1-z_2))\over q^2z_1z_2(1-q^2)(z_3-q^2z_2)}\nn\\
&+(z_1-1)(z_2-1){(q^2z_1-1)^2(1+q^2 z_2 z_3-z_2-z_3)\over z_2z_3(1-q^2)(z_2-q^2z_1)}
+{(q^2z_1-1)^2(z_3-1)^2\over z_2(1-q^2)}\nn\\
&-{(q^2z_1-1)^2(z_3-1)^3\over z_2(1-q^2)(z_3-q^2z_2)}+{(z_3-1)^2(q^2z_1-1)^3\over z_2(1-q^2)(z_2-q^2z_1)}
-{(z_3-1)^3(q^2z_1-1)^3\over z_2(1-q^2)(z_3-q^2z_2)(z_2-q^2z_1)}.
\end{align}
This is equivalent to
\begin{align}
&\sim\Bigl({(q^2z_1-1)^2(z_3-1)^2\over z_2(1-q^2)}
-{(q^2z_1-1)^2(z_3-1)^2-(q^2z_3-1)^2(z_1-1)^2q^2\over q^2z_1z_2(1-q^2)}\nn\\
&+{(q^2z_1-1)(z_1-1)((z_3-1)^2-(q^2z_3-1)^2)\over z_1(1-q^2)}+{(q^2z_3-1)(z_3-1)\over z_3(1-q^2)}
{(q^2z_1-1)^3-(z_2-1)^3\over z_2-q^2z_1}\Bigr)\nn\\
&+\Bigl(-{(q^2z_1-1)^3(z_3-1)^3\over q^2z_1z_2(1-q^2)(z_3-q^2z_2)}
+{(z_3-1)^3(q^2z_1-1)^3\over z_2z_3(1-q^2)(z_2-q^2z_1)}
-{(z_3-1)^3(q^2z_1-1)^3\over z_2(1-q^2)(z_3-q^2z_2)(z_2-q^2z_1)}\Bigr).\label{P3g}
\end{align}
We use following formula to simplify terms in the second parenthesis:
\begin{align}
&{(z_3-1)^3\over z_3-q^2z_2}f(z_1,z_2)\sim-{(q^2z_3-1)^3\over q^2(z_3-z_2)}f(z_1,z_2),\quad
{(q^2z_1-1)^3\over z_2-q^2z_1}f(z_2,z_3)\sim-{(z_1-1)^3\over q^2(z_2-z_1)}f(z_2,z_3).
\end{align}
The first parenthesis is simplified only by the exchange of variables. Then we have
\begin{align}
&{\rm (\ref{P3g})}\sim \Bigl({3q^2z_1^2z_3\over z_2}-3(1+q^2){z_1^2\over z_2}+8{z_1\over z_2}\Bigr)
+\Bigl({(q^2z_3-1)^3(q^2z_1-1)^3-q^4(z_3-1)^3(z_1-1)^3\over q^4(1-q^2)z_2z_3(z_2-z_1)}\nn\\
&-{(q^2z_3-1)^3(z_1-1)^3\over z_2q^2(1-q^2)(z_3-z_2)(z_2-z_1)}\Bigr).\label{P3h}
\end{align}
The last term in the second parenthesis is 
$$\sim-{(q^2z_3-1)^3(z_1-1)^3-(q^2z_1-1)^3(z_3-1)^3\over q^2(1-q^2)z_2(z_2-z_1)}.$$
The terms of the form $f(z_1,z_3)/(z_2(z_2-z_1))$ can be replaced by $-f(z_1,z_3)/(z_1z_2)+f(z_1,z_3)/(z_1(z_2-z_1))$. 
Then we have
\begin{align}
&{\rm (\ref{P3h})}\sim \Bigl({3q^2z_1^2z_3\over z_2}-3(1+q^2){z_1^2\over z_2}+8{z_1\over z_2}\Bigr)
-{(q^2z_3-1)^3(q^2z_1-1)^3-q^4(z_3-1)^3(z_1-1)^3\over q^4(1-q^2)z_1z_2z_3}\nn\\
&+{(q^2z_3-1)^2(z_1-1)^2+(q^2z_3-1)(z_3-1)(q^2z_1-1)(z_1-1)+(q^2z_1-1)^2(z_3-1)^2\over q^2z_1z_2}\nn\\
&+{1\over z_2-z_1}\Bigl((q^6+q^4+q^2+1)z_1^2z_3^2-3(q^4+q^2+1)z_1z_3(z_1+z_3)+3(q^2+1)(z_1^2+3z_1z_3+z_3^2)\nn\\
&-({z_1^2\over z_3}+{z_3^2\over z_1}+9(z_1+z_3))+{3\over q^2}({1\over z_1}+{1\over z_3})-{1+q^{-4}\over z_1z_3}\nn\\
&-{(q^2z_3-1)^2(z_1-1)^2+(q^2z_3-1)(z_3-1)(q^2z_1-1)(z_1-1)+(q^2z_1-1)^2(z_3-1)^2\over q^2z_1} \Bigr).\label{P3i}
\end{align}
Then all terms are in the form of $z_1^az_2^bz_3^c$ or $z_1^az_3^b/(z_2-z_1)$ with $2\ge a,b,c\ge -1$.
We should note the following relations
\begin{align}
&z_1^2f(z_2,z_3)\sim\Bigl({3(1+q^2)\over 1+q^4}z_1-{6\over 1+q^4}+{1+q^2\over q^2(1+q^4)z_1}\bigr)f(z_2,z_3),\nn\\&
{z_1^2\over z_2-z_1}f(z_3)\sim\Bigl({3\over q^2(z_2-z_1)}-{1+q^2\over q^4(z_2-z_1)z_1}-z_1\Bigr)f(z_3).
\end{align}
By these relations we can put powers $a,b,c$ are at $1,0,-1$. 
The canonical form is
\begin{align}
&{\rm (\ref{P3i})}\sim  \frac{(1+q^2)^2}{q^2}\frac{z_1}{z_3} 
+\frac{1}{(z_2-z_1)}\Bigl[\frac{12(1+q^2)}{ q^2} - \frac{(1+10q^2+q^4)z_1}{ q^2}
- \frac{3(1+q^2)^2}{q^4 z_1}  \nonumber \\
&+ z_3 \left\{ - \frac{1+10q^2+q^4}{q^2} + 3(1+q^2) z_1+ 
\frac{3(1+q^2)}{q^2 z_1} \right\}
+ \frac{1}{z_3} \left\{ - \frac{3(1+q^2)^2}{q^4} + \frac{3(1+q^2) z_1}{q^2}+ 
\frac{(1+q^2)(1+q^2+q^4)}{q^6 z_1} \right\}\Bigr].\label{P3d}
\end{align}
Multiple integral can be done through (\ref{B1}, \ref{B2}); 
\beqa
&&\displaystyle{i\over 8\phi^3}\int^{Q/2}_{-Q/2}{\rm d}^3 x {\bf W}_1(x_1+\epsilon_1){\bf W}_1(x_2+\epsilon_2)
{\bf W}_1(x_3+\epsilon_3){\rm(\ref{P3d})}= \displaystyle i\frac{e^{2i(\epsilon_3-\epsilon_1)\phi}}{16\phi^3} 
+{1\over 8\phi^3} \int^{\infty+i/2}_{-\infty+i/2}{{\rm d}x\over\sinh \pi x\sin\phi(x+\epsilon_1-\epsilon_2)} \nn\\
&&\displaystyle \times\Bigl[\frac{12(1+q^2)}{ q}e^{i\phi(\epsilon_1+\epsilon_2)}{\sin\phi x\over\sinh\phi} 
- \frac{(1+10q^2+q^4)}{ q^2}e^{i\phi(-x-\epsilon_1+\epsilon_2)}x 
- \frac{3(1+q^2)^2}{q^2 }e^{i\phi(x+3\epsilon_1+\epsilon_2)}{\sin 2\phi x\over\sinh 2\phi} \nn\\  
&&\displaystyle+ {e^{-2i\phi\epsilon_3}\over\cosh\phi} \Bigl\{ - \frac{1+10q^2+q^4}{q^2}
e^{i\phi(\epsilon_1+\epsilon_2)}{\sin\phi x\over\sinh\phi}+ \frac{3(1+q^2)}{q}
e^{i\phi(-x-\epsilon_1+\epsilon_2)}x + 
\frac{3(1+q^2)}{q } e^{i\phi(x+3\epsilon_1+\epsilon_2)}{\sin 2\phi x\over\sinh 2\phi}\Bigr\} \nn \\
&&\displaystyle+ {e^{2i\phi\epsilon_3}\over\cosh\phi} \Bigl\{ - \frac{3(1+q^2)^2}{q^2}
e^{i\phi(\epsilon_1+\epsilon_2)}{\sin\phi x\over\sinh\phi}  + \frac{3(1+q^2)}{q}
e^{i\phi(-x-\epsilon_1+\epsilon_2)}x + 
\displaystyle\frac{(1+q^2)(1+q^2+q^4)}{q^3}e^{i\phi(x+3\epsilon_1+\epsilon_2)}{\sin 2\phi x\over\sinh 2\phi}
 \Bigr\}\Bigr].\nn\\ \label{P3f}
\enqa
Then we have
\begin{align}
P(3)={1\over 2} \lim_{\epsilon_1,\epsilon_2,\epsilon_3\to 0}(\partial_2-\partial_1)(\partial_3-\partial_1)
(\partial_3-\partial_2)
{\rm (\ref{P3f})}={\rm (\ref{P3new})}.
\end{align}

Equation (\ref{transverse2}) is
\begin{align}
\langle S_{j}^x S_{j+2}^x \rangle &= {i\over 8\phi^3}
\int_{-Q/2}^{Q/2} {\rm d}^3 x{\bf W}_3
\frac{(q^2z_1-1)^2(q^2z_2-1)(z_2 -1) 
(z_3 -1)^2}{q^2(z_2-z_1)(z_3-z_1) 
(z_3-q^2z_2)}.
\label{transverse2b}
\end{align}
The integrand is devided as follows
\begin{align}
{(q^2z_1-1)^2(q^2z_2-1)(z_2 -1)(z_3 -1)^2\over q^2(1-q^2)z_2}
\Bigl( {1\over (z_3-z_1)(z_3-q^2z_2)}
+{1\over (z_3-z_1)(z_2-z_1)}
-{1\over (z_3-q^2z_2)(z_2-z_1)}\Bigr).\label{tra2c}
\end{align}
The first term is transformed as \begin{align}
&{(q^2z_1-1)^2(q^2z_2-1)(z_2 -1)(z_3 -1)^2\over q^2(1-q^2)z_2(z_3-z_1)(z_3-q^2z_2)}
={(q^2z_1-1)^2(z_2 -1)(q^2z_2-1)\over q^2(1-q^2)z_2(z_3-z_1)}(z_3+q^2z_2-2+{(q^2z_2-1)^2\over z_3-q^2z_2})\nn\\
&\sim{(q^2z_1-1)^2(z_2 -1)(q^2z_2-1)\over q^2(1-q^2)z_2(z_3-z_1)}(z_3+q^2z_2-2)-
{(q^2z_1-1)^2(q^{-2}z_2 -1)(z_2-1)^3\over (1-q^2)z_2(z_3-z_1)(z_3-z_2)}.
\end{align}
The third term is transformed as
\begin{align}
&-{(q^2z_1-1)^2(q^2z_2-1)(z_2 -1)(z_3 -1)^2\over q^2(1-q^2)z_2(z_3-q^2z_2)(z_2-z_1)}=
-{(q^2z_1-1)^2(z_2 -1)(z_3 -1)^2\over q^2(1-q^2)z_2(z_2-z_1)}(-1+{z_3-1\over z_3-q^2z_2})\nn\\
&\sim{(q^2z_1-1)^2(z_2 -1)(z_3 -1)^2\over q^2(1-q^2)z_2(z_2-z_1)}+{(q^2z_1-1)^2(z_2 -1)(q^2z_3 -1)^3
\over q^4(1-q^2)z_2(z_2-z_1)(z_3-z_2)}.
\end{align}
Using the equivalence
\begin{align}
{f(z_1,z_2,z_3)\over(z_1-z_2)(z_1-z_3)}\sim {1\over z_1-z_2}\Bigl({f(z_1,z_2,z_3)-f(z_1,z_3,z_2)\over z_2-z_3}\Bigr),
\end{align}
we replace the term like $(z_1-z_2)(z_1-z_3)$ in the denominator by $z_1-z_2$. 
The canonical form is
\begin{align}
&{\rm (\ref{tra2c})}\sim \frac{1}{z_2-z_1}\Bigl[- \frac{(1+q^2)(3+q^4)}{q^4} + \frac{(3- 2q^2+3 q^4) z_1}{q^2} 
+ \frac{(1+q^2)^2}{ q^6 z_1} \nonumber \\
&+ z_3 \left\{ \frac{6-2q^2}{q^2} - \frac{(1+q^2) z_1}{q^2}+ \frac{(1+q^2)(-2+q^2)}{q^4 z_1} \right\} 
+ \frac{1}{z_3} \left\{\frac{4+4q^2}{q^2} - (1+q^2) z_1- \frac{1+q^2}{q^4 z_1} \right\}\Bigr]. \label{tra2d}
\end{align}
Then the integral of this form is 
\begin{align}
&{i\over 8\phi^3}\int^{Q/2}_{-Q/2}{\rm d}^3 x {\bf W}_1(x_1+\epsilon_1){\bf W}_1(x_2+\epsilon_2)
{\bf W}_1(x_3+\epsilon_3){\rm(\ref{tra2d})}\nn\\
&={1\over 8\phi^3} \int^{\infty+i/2}_{-\infty+i/2}{{\rm d}x\over\sinh \pi x\sin\phi(x+\epsilon_1-\epsilon_2)} \nn\\
&\times\Bigl[- \frac{(1+q^2)(3+q^4)}{q^3}e^{i\phi(\epsilon_1+\epsilon_2)}{\sin\phi x\over\sinh\phi}  
+ \frac{(3- 2q^2+3 q^4)}{q^2} e^{i\phi(-x-\epsilon_1+\epsilon_2)}x 
+ \frac{(1+q^2)^2}{ q^2}e^{i\phi(x+3\epsilon_1+\epsilon_2)}{\sin 2\phi x\over\sinh 2\phi}
 \nonumber \\
&+ {e^{-2i\phi\epsilon_3}\over\cosh\phi} \left\{ \frac{6-2q^2}{q}e^{i\phi(\epsilon_1+\epsilon_2)}
{\sin\phi x\over\sinh\phi}  - \frac{(1+q^2)}{q}e^{i\phi(-x-\epsilon_1+\epsilon_2)}x + 
\frac{(1+q^2)(-2+q^2)}{q^2}e^{i\phi(x+3\epsilon_1+\epsilon_2)}{\sin 2\phi x\over\sinh 2\phi}
 \right\} 
\nonumber \\
&+ {e^{2i\phi\epsilon_3}\over\cosh\phi} \left\{\frac{4+4q^2}{q}e^{i\phi(\epsilon_1+\epsilon_2)}
{\sin\phi x\over\sinh\phi}  - \frac{1+q^2}{q}e^{i\phi(-x-\epsilon_1+\epsilon_2)}x 
- \frac{1+q^2}{q}e^{i\phi(x+3\epsilon_1+\epsilon_2)}{\sin 2\phi x\over\sinh 2\phi}
 \right\}\Bigr].\label{tra2e}
\end{align}
Then we have
\begin{align}
&\langle S_{j}^x S_{j+2}^x \rangle={1\over 2} \lim_{\epsilon_1,\epsilon_2,\epsilon_3\to 0}
(\partial_2-\partial_1)(\partial_3-\partial_1)(\partial_3-\partial_2)
{\rm (\ref{tra2e})}={\rm (\ref{transnew})}.
\end{align}
\end{widetext}

%

\end{document}